\documentclass[10pt,letterpaper,compsoc,conference]{iiswc26}

\usepackage{cite}
\usepackage{amsmath,amssymb,amsfonts}
\usepackage{algorithmic}
\usepackage{algorithm}
\usepackage{graphicx}
\usepackage[dvipsnames]{xcolor}
\usepackage[final]{microtype}
\usepackage[italic]{mathastext}
\usepackage{libertine}
\usepackage[T1]{fontenc}
\usepackage{textcomp}
\usepackage[varqu,varl]{zi4}
\usepackage[all]{nowidow}
\usepackage[keeplastbox]{flushend}
\usepackage{fancyhdr}
\usepackage{url}
\usepackage[most]{tcolorbox}
\usepackage{booktabs}

\makeatletter
\g@addto@macro\UrlBreaks{\UrlOrds}
\makeatother

\newcounter{observation}
\newtcolorbox{observationbox}[1][]{%
  enhanced, breakable,
  before skip=4pt, after skip=4pt,
  colback=white, colframe=black,
  boxrule=0.4pt, arc=2pt,
  left=4pt, right=4pt, top=3pt, bottom=3pt,
  #1
}
\newenvironment{observation}[1][]{%
  \refstepcounter{observation}%
  \begin{observationbox}[#1]\itshape\noindent
  \textbf{Observation~\#\theobservation:}\ \ignorespaces%
}{\end{observationbox}}

\begin{document}

\title{When and Why to Use TMA: A Workload-Grounded Characterization of
Tensor Memory Acceleration on NVIDIA Hopper GPUs}

\author{\IEEEauthorblockN{Christin David Bose, Cesar Avalos Baddouh, Junrui Pan, Timothy G. Rogers}
\IEEEauthorblockA{Purdue University, West Lafayette, IN, USA\\
\{chris241, cavalosb, pan251, timrogers\}@purdue.edu}}

\maketitle
% No page numbers and no headers/footers of any type, per the
% formatting requirements. (The shipped template uses \pagestyle{plain},
% which numbers every page from 2 onward.)
\pagestyle{empty}
\thispagestyle{empty}

% Activates the @IEEEtranBSTCTL entry in reference.bib (no dash for a
% repeated author name). Produces no visible output.
\bstctlcite{BSTcontrol}

\begin{abstract}
The Tensor Memory Accelerator (TMA), introduced on NVIDIA Hopper,
is used pervasively in production GPU kernels yet its usage is
conspicuously selective: expert kernel developers in CUTLASS,
FlashAttention-3, and FlashInfer apply TMA to some tensor operands
within a kernel but fall back to cp.async, direct global loads, or
distributed shared memory (DSM) for others. Vendor documentation
describes the API, and recent microbenchmarking work characterizes
TMA throughput in isolation, but neither answers the question
practitioners actually face: given a memory operand in a kernel,
should it use TMA, and if so, with or without cluster multicast?
We address this gap with a workload-grounded characterization study
on H100. We catalog the TMA, multicast, DSM, and cp.async choices
made inside FlashAttention-3 forward, isolate the discriminating
features through targeted microbenchmarks (transfer-size crossover,
multicast scaling), and produce rules of thumb for the
load-mechanism choice on Hopper. We
further compare TMA multicast against DSM as competing
cluster-broadcast primitives, and isolate the FlashAttention-3
paged-key-value (KV) TMA-vs-cp.async dispatch on H100 across page size,
datatype, model shape, and execution phase. Our results yield concrete,
predictive guidance on when TMA pays off and when multicast helps
or hurts on Hopper, and we extend the analysis to server Blackwell
(SM100) via a CUTLASS GEMM study on B200 that shows the qualitative
dispatch rules carry across generations while absolute crossovers
shift with tensor-core and bandwidth scaling.
\end{abstract}

\section{Introduction}
\label{sec:intro}
\begin{figure}[t]
  \centering
  \includegraphics[width=\columnwidth]{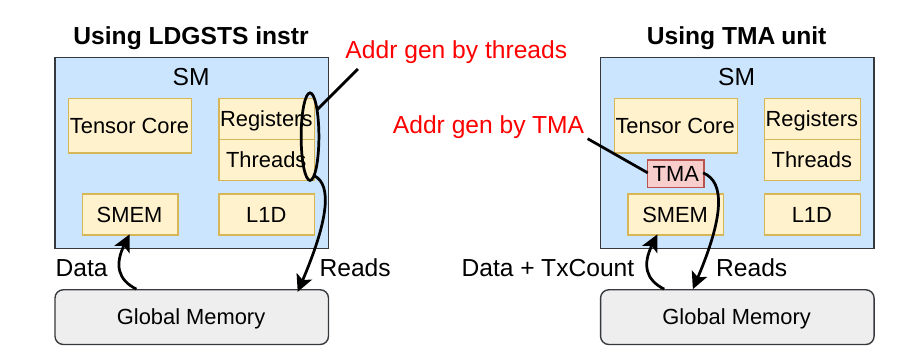}
  \caption{High-level view of the \texttt{cp.async} and TMA paths on Hopper}
  \label{fig:tma_schematic}
\end{figure}

The Tensor Memory Accelerator (TMA) is the headline memory-system
addition of NVIDIA's Hopper architecture~\cite{nvidia_hopper_h100}.
By issuing a single hardware-managed bulk asynchronous copy from
global to shared memory, indexed through a precomputed tensor
descriptor, TMA frees warps from generating per-thread addresses
and overlaps data movement with tensor-core compute
(Figure~\ref{fig:tma_schematic}). It also exposes two features
that have no equivalent in the prior \texttt{cp.async} 
path: hardware-supported swizzling for bank-conflict-free SMEM
layouts feeding tensor cores, and a \emph{multicast} mode that broadcasts
the same tile to multiple SMs in a thread-block cluster with a
single issue and a single L2 read.

TMA is used heavily in production. CUTLASS~\cite{cutlass} dispatches
its SM90 GEMM mainloops through TMA for the operand matrices;
FlashAttention-3~\cite{flashattention3} routes Q, K, and V through
TMA on its prefill path; FlashInfer~\cite{flashinfer} uses TMA for
contiguous KV. Public write-ups attribute much of the resulting SM90 uplift
to TMA~\cite{colfax_tma_tutorial,pytorch_tma_fp8}.

\textbf{And yet TMA is not used for everything.} Expert kernel code
selectively avoids TMA in regimes where one might naively expect it
to win. Three concrete examples motivate this paper.

\emph{First}, FlashAttention-3 contains an in-tree dispatch
heuristic that disables the TMA-paged-KV path when per-tile work
falls below a compute-bound threshold~\cite{flashattention3_repo}.
The heuristic was shipped on empirical grounds alone, without an
articulated mechanism. A companion
issue~\cite{fa3_issue_1572} reports a related TMA overhead pattern
in deployed batch decode.

\emph{Second}, FlashInfer~\cite{flashinfer} explicitly states that
TMA is used \emph{only} for contiguous KV on Hopper, and falls back
to \texttt{cp.async} for paged KV. PagedAttention~\cite{pagedattention}
and vAttention~\cite{vattention} treat the KV page-size knob as a
purely allocator-level tradeoff (small pages waste fewer tokens at
the tail; large pages amortize lookup cost), but the choice of page
size also dictates whether TMA is even usable for the KV
load (FA3 requires the page size to be a multiple of the kernel's
K-tile), and operators of
vLLM~\cite{vllm} and TensorRT-LLM~\cite{trtllm} currently guess the
page size without quantitative guidance on how much TMA bandwidth
they leave on the cp.async fallback.

\emph{Third}, the choice between \emph{TMA multicast} and
\emph{distributed shared memory} (DSM), the two hardware mechanisms
Hopper provides for cluster-wide tile sharing, is unsettled even
among expert codebases. CUTLASS GEMMs broadcast operand tiles
across cluster CTAs via TMA multicast~\cite{cutlass}, while
community reports~\cite{forum_multicast_slow,forum_tma_oddity}
flag regimes in which multicast underperforms independent unicast
TMAs from the same cluster. Both hardware paths solve the same
logical problem (get one Global Memory (GMEM) tile to N SMs in a cluster), but no
published work characterizes when one beats the other, and DSM as
an alternative cluster-broadcast primitive has received no
systematic comparison at all.

These three observations are not independent oddities; they share a
common cause. TMA optimizes for a specific access regime (large,
multi-dimensional, well-aligned tiles whose descriptor cost amortizes
over many issues) and degrades, sometimes sharply, outside that
regime. Production code encodes this fact in dispatch heuristics
written by experts, but the heuristics are workload-specific,
under-documented, and not always articulated by their
authors. NVIDIA's
Hopper Tuning Guide~\cite{nvidia_hopper_tuning} describes the API
but does not provide crossover points; recent academic
microbenchmarking work characterizes TMA throughput and latency in
isolation, or the cp.async path in isolation, but does not address
the comparative question~\cite{luo_hopper,luo_hopper_arxiv,jarmusch_blackwell,luhnen_thesis,li_asyncmemcpy,li_taco_pipelining}.

\textbf{The questions this paper answers.} We frame the gap as three
concrete questions a kernel developer can act on:

\begin{enumerate}
  \item \emph{When does TMA outperform \texttt{cp.async}, and by how
    much?} What are the crossover points in transfer size and
    bytes-in-flight per SM, and how do they manifest in production
    attention across page size, datatype, model shape, and execution
    phase?
  \item \emph{When does TMA multicast pay off, and when does it
    hurt?} How does the answer depend on cluster size and tile size,
    and is the L2 Request Coalescer (LRC)-deduplicated redundant-unicast path a better
    choice in the streaming regime?
  \item \emph{Which cluster-broadcast primitive should one pick?}
    How does TMA multicast compare head-to-head against
    distributed-shared-memory (DSM) broadcast when the DRAM cost is
    matched?
\end{enumerate}

\textbf{Approach.} We answer these questions through a
workload-grounded characterization study. We first \emph{catalog} the
memory-mechanism choice (TMA, TMA multicast, DSM, \texttt{cp.async}) for every operand of one production kernel:
FlashAttention-3 forward, on prefill, chunked prefill, and decode
paths across dense multi-head attention (MHA), grouped-query
attention (GQA), and multi-head latent attention (MLA) shapes.
From the catalog we form \emph{hypotheses} for the discriminating
features (dimensionality, transfer size, reuse, multicast
applicability, multicast-vs-DSM preconditions). We then test these hypotheses with targeted
microbenchmarks on H100 (SM90) that sweep the relevant axes in
isolation and measure both wall-clock time and the underlying
GPU metrics with NVIDIA Nsight Compute
(NCU)~\cite{ncu_guide}.

\textbf{A cross-cutting case study} runs through the research
questions: the FlashAttention-3~\cite{flashattention3} paged-KV
TMA-vs-\texttt{cp.async} dispatch on H100. FA3's modular dispatch
means the TMA-paged path engages only at page sizes that are
multiples of the kernel's K-tile, while every other commonly
deployed page size falls back to \texttt{cp.async}, a
configuration whose performance impact operator-facing tools have
not yet measured. Sweeping page size across dense MHA
(Llama-3-8B~\cite{llama3}), GQA (Llama-3-70B), and MLA
(DeepSeek-V2~\cite{deepseekv2}, with FlashMLA~\cite{flashmla} as
the production reference) directly answers a question
vLLM/SGLang/TensorRT-LLM operators currently guess at, and
isolates whether FA3's compute-bound gate is empirically necessary
across the deployed model.

\textbf{Contributions.} We make three contributions, mapped
one-to-one to the research questions above:

\begin{itemize}
  \item \emph{(Q1)} A workload-grounded characterization of when
    TMA outperforms \texttt{cp.async} on Hopper, combining a
    microbench crossover study with a controlled sweep comparing
    TMA-paged against forced
    \texttt{cp.async}-paged on FA3 across page size, datatype, model
    shape, and execution phase.
  \item \emph{(Q2)} A characterization of TMA multicast against
    the LRC-deduplicated redundant-unicast path as competing
    intra-cluster fan-out primitives, swept across cluster size
    and tile size, identifying the regime in which redundant
    unicast is faster end-to-end and the regime in which multicast's L2 cleanliness is
    essential.
  \item \emph{(Q3)} A head-to-head comparison of TMA multicast
    against DSM broadcast at matched
    DRAM cost, across cluster size and tile size on H100, showing
    when each primitive is the right choice for the regular
    broadcast pattern.
\end{itemize}

The remainder of the paper is organized as follows. Section~II
provides background on TMA, \texttt{cp.async}, thread-block
clusters, multicast, and DSM on Hopper. Section~III describes the
experimental methodology and workload corpus. Section~IV reports
the characterization results, including the FA3 paged-KV
TMA-vs-\texttt{cp.async} case study and the multicast-vs-redundant-unicast
comparison. Section~V extends the characterization to server Blackwell (SM100)
via a CUTLASS GEMM study on B200. Section~VI discusses related work.
Section~VII concludes.

\section{Background}
\label{sec:background}

This section reviews the GPU memory mechanisms that this paper
characterizes and compares. We focus on the level of detail required
to interpret the experiments in Section~\ref{sec:evaluation}. We assume basic familiarity with the CUDA
execution model.

\subsection{The asynchronous-pipeline ideal on Hopper}

The performance ceiling of a Hopper tensor-core kernel is set by
how completely the warp groups doing math overlap the warps doing
memory movement. CUTLASS~\cite{cutlass}, FlashAttention-3
\cite{flashattention3}, and FlashInfer~\cite{flashinfer} all build
on the same template: one or more \emph{producer} warps stream tiles
from global memory (GMEM) into shared memory (SMEM) in software
pipelined stages, and one or more \emph{consumer} warp groups issue
asynchronous Warp Group MMA (WGMMA) tensor-core operations
out of those SMEM stages. Producers and consumers synchronize
through \texttt{mbarrier} objects in SMEM. The kernel hits peak when
the producer can keep \(N\) tile stages full while the consumers
drain them at WGMMA throughput.

Two features of Hopper directly raise this ceiling: the Tensor
Memory Accelerator and thread-block clusters. The choice of
\emph{which} memory primitive the producer warps use to fill SMEM
(TMA, TMA with multicast, distributed shared memory (DSM), or the
older \texttt{cp.async} (LDGSTS) path) is the choice this paper
characterizes. Algorithms~\ref{alg:cpasync_producer_consumer}
and~\ref{alg:tma_producer_consumer} sketch the same
producer/consumer pipeline filled by \texttt{cp.async} versus TMA:
the TMA variant replaces every producer thread's address
arithmetic and per-thread issue stream with a single
\texttt{cp.async.bulk.tensor} from one elected thread against a
precomputed tensor-map descriptor.

\newcommand{\cpa}[1]{\textcolor{blue}{\texttt{#1}}}
\begin{algorithm}[t]
\small
\caption{\texttt{cp.async} enabled producer/consumer pipeline.}
\label{alg:cpasync_producer_consumer}
\begin{algorithmic}[1]
\IF{producer warpgroup}
  \FOR{$\text{tile} \in \text{tiles}$}
    \STATE $b \gets \text{tile} \bmod 2$
    \STATE \texttt{wait\_until\_buffer\_free}($b$)
    \STATE $\texttt{gmem\_ptr} \gets \texttt{\&global[tile][thread\_slice]}$
    \STATE $\texttt{smem\_ptr} \gets \texttt{\&smem[}b\texttt{][thread\_slice]}$
    \STATE \cpa{cp.async.shared.global}(\texttt{smem\_ptr}, \texttt{gmem\_ptr}, $B_{\text{th}}$)
    \STATE \cpa{cp.async.commit\_group}()
    \STATE \cpa{cp.async.wait\_group}(0)
    \STATE \texttt{signal\_buffer\_ready}($b$)
  \ENDFOR
\ENDIF
\IF{consumer warpgroup}
  \FOR{$\text{tile} \in \text{tiles}$}
    \STATE $b \gets \text{tile} \bmod 2$
    \STATE \texttt{wait\_until\_buffer\_ready}($b$)
    \STATE \texttt{compute\_on\_tile}(\texttt{smem[}$b$\texttt{]})
    \STATE \texttt{signal\_buffer\_free}($b$)
  \ENDFOR
\ENDIF
\end{algorithmic}
\end{algorithm}

\begin{algorithm}[t]
\small
\caption{TMA enabled producer/consumer pipeline.}
\label{alg:tma_producer_consumer}
\begin{algorithmic}[1]
\IF{producer warpgroup}
  \FOR{$\text{tile} \in \text{tiles}$}
    \STATE $b \gets \text{tile} \bmod 2$
    \STATE \texttt{wait\_until\_buffer\_free}($b$)
    \STATE $\texttt{coords} \gets \texttt{tensor\_tile\_coords(tile)}$
    \STATE \cpa{tma.load}(\texttt{dst\_smem}=\texttt{smem[}$b$\texttt{]},
              \texttt{tensor\_map}=$\texttt{desc}$,
              \texttt{coords}=\texttt{coords},
              \texttt{completion\_barrier}=\texttt{ready[}$b$\texttt{]})
  \ENDFOR
\ENDIF
\IF{consumer warpgroup}
  \FOR{$\text{tile} \in \text{tiles}$}
    \STATE $b \gets \text{tile} \bmod 2$
    \STATE \texttt{wait\_until\_buffer\_ready}($b$)
    \STATE \texttt{compute\_on\_tile}(\texttt{smem[}$b$\texttt{]})
    \STATE \texttt{signal\_buffer\_free}($b$)
  \ENDFOR
\ENDIF
\end{algorithmic}
\end{algorithm}

\begin{comment}
\begin{figure}[t]
  \centering
  \includegraphics[width=\columnwidth]{figs/TMA_algo.pdf}
  \caption{Producer-side issue model for the two pre-cluster
  GMEM$\to$SMEM paths. With \texttt{cp.async} (left), every thread
  in the warp computes its own GMEM and SMEM addresses, issues an
  LDGSTS, and the warp synchronizes via a
  \texttt{commit\_group}/\texttt{wait\_group} pair before the tile
  is consumed. With TMA (right), a single elected thread issues one
  \texttt{cp.async.bulk.tensor} against a precomputed
  \texttt{CUtensorMap}, and the TMA engine walks the descriptor,
  applies the hardware swizzle, and signals an \texttt{mbarrier}
  when the bulk transfer completes; the other threads in the warp
  do no address arithmetic.}
  \label{fig:tma_algo}
\end{figure}
\end{comment}

Algorithms~\ref{alg:cpasync_producer_consumer} and~\ref{alg:tma_producer_consumer} contrast the two paths at the
producer-warp level. The structural difference is where address
generation lives: \texttt{cp.async} pushes per-thread address
arithmetic into the warp's issue stream every iteration, while TMA
hoists the address computation into a one-time descriptor build and
reduces the per-iteration cost to a single bulk-issue plus a
barrier wait. This distinction (per-thread vs.\ single-thread issue,
per-tile vs.\ per-tensor address logic) is the structural source of
every quantitative crossover we report in
Section~\ref{sec:evaluation}.

\subsection{\texttt{cp.async} (LDGSTS): the pre-Hopper baseline}

\texttt{cp.async}, exposed at the SASS level as
LDGSTS, was introduced on Ampere (SM80)~\cite{nvidia_a100} and
remains the fallback path on Hopper. Each thread
issues its own load with a per-thread address. The instruction is
asynchronous in that the issuing thread does not stall on the load; completion
is established with an explicit \texttt{cp.async.wait\_group} (paired with a
CTA barrier) before the SMEM contents are consumed. \texttt{cp.async} loads
go through the L1/tex pipe and SMEM is filled per-thread.

Two properties of \texttt{cp.async} matter for the comparison:
(i) the address generation cost scales linearly with the number of
participating threads, so issuing a tile load consumes warp
issue-slot bandwidth and burns instruction-cache pressure; and
(ii) because each thread writes its own SMEM bytes, achieving
bank-conflict-free SMEM layouts for downstream WGMMA requires
\emph{software} swizzling in the address calculation. Li et
al.~\cite{li_asyncmemcpy,li_taco_pipelining} characterize the
\texttt{cp.async} (Async Memcpy) path itself on A100 and H100,
reporting the regimes in which it overlaps usefully with compute;
the Hopper microbenchmarking work of Luo et
al.~\cite{luo_hopper_arxiv} reports the per-instruction overhead and
L1-pipe behavior in detail. Neither line of work compares
\texttt{cp.async} against TMA on the same operand, which is the
comparison this paper makes.

\subsection{Tensor Memory Accelerator (TMA)}

TMA, exposed in PTX as the \texttt{cp.async.bulk.tensor} family~\cite{nvidia_ptx_isa},
replaces the per-thread address-generation model of \texttt{cp.async}
with a hardware-managed bulk transfer. A single thread in a CTA
issues one TMA instruction, and a dedicated hardware unit (the TMA
engine, one per SM) walks a precomputed tensor descriptor and
streams a multi-dimensional tile from GMEM into SMEM. Three aspects
of the design drive the comparisons in this paper.

\textbf{Tensor descriptors.} A TMA load is parameterized by a
\texttt{CUtensorMap} object built on the host (or on-device via
\texttt{tensormap.replace}) that encodes the GMEM base pointer,
per-dimension stride and extent, tile shape, fill mode, and swizzle
pattern. The descriptor is shared across all TMA issues that index
into the same logical tensor. Descriptor construction is a
one-time per-tensor setup performed outside the kernel's inner
loops; each TMA \emph{issue} against a built descriptor then adds
a $\sim$170-cycle per-access overhead before the bulk transfer
begins~\cite{luo_hopper_arxiv}. The descriptor cost is therefore a fixed
overhead amortized over all subsequent issues against the same
tensor, and the per-issue cost is amortized over the bulk transfer
size: tiny transfers or low descriptor reuse leave both costs
unamortized.

\textbf{Shape constraints.} A TMA tensor descriptor supports 1-D
through 5-D access. Each non-stride dimension is capped at
256~elements, which makes 1-D TMA practically limited to about
2\,KB per issue~\cite{luo_hopper_arxiv}. Higher-rank descriptors lift
the per-issue size but introduce per-rank stride and box-shape
constraints; in particular, the descriptor's leading (\emph{x})
dimension extracts higher throughput than the trailing (\emph{y},
\emph{z}) dimensions, an asymmetry first reported in
\cite{luo_hopper_arxiv}. TMA addresses
must be 16-byte aligned and the descriptor stride along the inner
dimension must be a multiple of 16 bytes; misalignment is a hard
correctness requirement, not a performance penalty.

\textbf{SMEM path and hardware swizzle.} TMA bypasses L1 entirely,
reads through L2, and writes directly to SMEM~\cite{luo_hopper_arxiv}.
The descriptor specifies one of four hardware swizzle patterns
(none, 32B, 64B, 128B) which the TMA engine applies during the
write so that the resulting SMEM layout is bank-conflict-free for
WGMMA consumption. 

\textbf{Completion.} TMA completion is signaled through a synchronization primitive such as an \texttt{mbarrier}~\cite{nvidia_ptx_mbarrier} in SMEM that the issuing thread arms with the
expected byte count; consumer warps wait on the barrier before
reading the tile. The same mbarrier infrastructure is shared with
tensor core accumulator tracking, which is what makes the
producer/consumer pipeline efficient.

\subsection{Thread-block clusters}

\begin{figure}[t]
\centering
\includegraphics[width=\columnwidth]{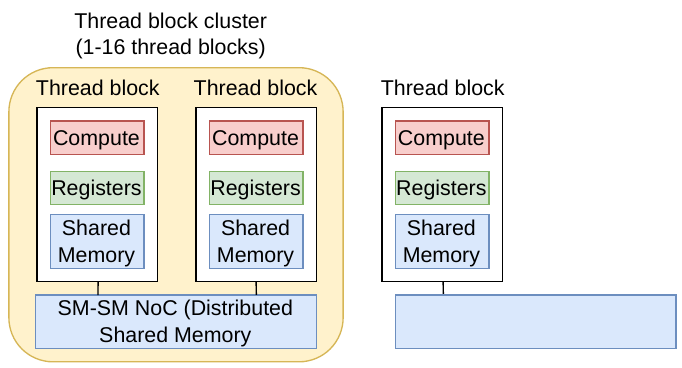}
\caption{Illustration of a thread block cluster}
\label{fig:TBC}
\end{figure}

Hopper introduced the thread-block cluster (Figure~\ref{fig:TBC}), a new grouping above
the Cooperative Thread Array (CTA) in the execution hierarchy~\cite{nvidia_hopper_h100}. A
cluster is a set of up to 16 CTAs
co-scheduled onto adjacent SMs in a Graphics Processing Cluster
(GPC). Cluster CTAs can synchronize through cluster-wide barriers,
and, critically for this paper, they can address each other's
SMEM via the \emph{Distributed Shared Memory} (DSM) Network-on-Chip (NoC)
described below. Clusters are the unit on which both TMA multicast
and DSM operate; outside a cluster, neither mechanism is available. NVIDIA H100~\cite{nvidia_hopper_h100} GPUs have also introduced an L2 Request Coalescer (LRC) that processes incoming requests for L2 and tries to coalesce read requests before forwarding them to the L2 cache~\cite{ncu_guide}.

\subsection{TMA multicast}

A TMA \texttt{cp.async.bulk.tensor}~\cite{nvidia_ptx_isa} instruction can carry a
\emph{multicast mask}: a bitmap indicating which CTAs in the
issuing cluster should receive the loaded tile. The hardware
performs a single GMEM read, a single L2 lookup, and writes the
tile into the SMEM of every masked CTA. Each receiving CTA ends up
with its own complete SMEM copy, so subsequent reads are local-SMEM
fast.

Multicast is the standard cluster-broadcast primitive in CUTLASS
GEMM~\cite{cutlass}: when the cluster is partitioned along the
M~dimension and all CTAs need the same B~tile (or vice versa),
multicast reduces the GMEM/L2 traffic by a factor equal to the
multicast degree. Two operational details matter for the
characterization. First, the multicast issue must come from one
elected leader CTA; the other receivers wait on an mbarrier.
Second, multicast still costs SMEM proportional to the multicast
degree (one full copy per receiver), which competes with the SMEM
budget available for pipeline stages. Community reports~\cite{forum_multicast_slow} of multicast
underperforming independent unicast TMAs are consistent with the
issue serializing on the leader and with the L2 already absorbing
duplicate reads in the non-multicast case; Section~\ref{sec:evaluation} isolates the
regimes in which this holds.

\subsection{Distributed shared memory (DSM)}

DSM exposes the SMEM of every CTA in a cluster as a single
distributed address space~\cite{nvidia_hopper_h100,luhnen_thesis}.
A thread in CTA~\(i\) can issue an ordinary SMEM load or store
whose address falls inside CTA~\(j\)'s SMEM partition; the request
travels over the cluster crossbar to the remote SM and back.
Latency is higher than local SMEM access but substantially lower
than going through L2~\cite{luo_hopper_arxiv,luhnen_thesis}.

DSM and TMA multicast solve the same logical problem (getting one
GMEM tile to \(N\) cluster SMs) through different hardware paths.
With DSM, one CTA loads the tile into its local SMEM (typically via
TMA), and sibling CTAs read the tile remotely over the crossbar;
only one SMEM copy exists. With multicast, every receiving CTA
holds its own copy. The tradeoff is therefore SMEM capacity (DSM
wins, by a factor of the cluster size) against per-access latency
and bandwidth (DSM loses on each remote read, but the loss is
constant per access), and the crossover depends on the
\emph{reuse count}: how many times each consumer re-reads the tile.
CUTLASS~\cite{cutlass} chose multicast for operand broadcast in its
GEMM mainloops, where reuse is high; production attention kernels
have at times reached for DSM in low-reuse, SMEM-constrained
regimes. Section~\ref{sec:evaluation} quantifies when each choice is correct.

\section{Experimental Methodology}
\label{sec:method}

This section describes the hardware platforms, software stack,
workload corpus, microbenchmark harness, and measurement protocol
used to produce the results in Section~\ref{sec:evaluation}. 

\subsection{Hardware platforms}

We evaluate on NVIDIA H100 SXM5 (SM90, Hopper): 132 SMs, 50\,MB L2,
80\,GB HBM3 at 3.35\,TB/s peak, 228\,KB SMEM per SM, cluster size
up to 16 CTAs, single TMA engine per
SM~\cite{nvidia_hopper_h100}. Each measurement is
preceded by a warmup that reaches a steady thermal state.
Section~\ref{sec:blackwell} extends the CUTLASS GEMM sweep to
server Blackwell (B200, SM100); the H100 setup above is otherwise
held constant.

\subsection{Software stack}

All measurements use CUDA Toolkit 12.6 with driver 560.35 on H100 SXM5. The production
kernel in
the catalog is taken from the upstream
FlashAttention-3~\cite{flashattention3_repo} repository.

\subsection{Workload corpus}

\textbf{FlashAttention-3}~\cite{flashattention3_repo} forward,
prefill and decode paths, is the production kernel we
evaluate end-to-end. FA3 routes Q, K, and V through TMA on the
contiguous-KV path, falls back to \texttt{cp.async} for paged KV,
and gates the TMA-paged path on an in-tree compute-bound predicate.
We extract every GMEM\(\to\)SMEM load site in FA3, label the
mechanism actually issued at runtime (TMA, TMA multicast,
\texttt{cp.async}), and record the
source-level conditions that gate the dispatch. Mechanism labels
are cross-checked against SASS at the JIT-symbol level for the FA3
paged-KV dispatch, where the source-level predicate is non-trivial.

\textbf{CUTLASS GEMM} mainloops on SM90 dispatch TMA (and TMA multicast
for operand broadcast) by default; they let us test the same
TMA-vs-\texttt{cp.async} question on a non-attention workload. We
benchmark 18 production GEMM shapes drawn from the projection layers
of Llama-3-8B, Llama-3-70B, and DeepSeek-V2 (both dense and MLA
compression projections). Per shape and datatype we sweep the dimension \(M \in \{16, 64, 256, 512, 2048, 8192\}\), giving \(18 \times 6
\times 3 = 324\) cells; each cell is a comparison between the
TMA-dispatched CUTLASS kernel and a forced \texttt{cp.async}-only
variant of the same tile schedule. For each TMA/cp.async variant, we choose the tile size that achieves the lowest runtime. Wall-time is the median of 100
timed iterations; run-to-run variation stays below 1\%.

\textbf{Workloads considered but not measured.} We also examined
FlashInfer~\cite{flashinfer_repo} and its FlashMLA~\cite{flashmla}
companion as candidate workloads, but both currently dispatch only
\texttt{cp.async} on the paged-KV path; enabling TMA-paged for an
apples-to-apples comparison would require a significant rewrite of
each kernel's producer warpgroup, which is beyond the scope of this
characterization study.

Table~\ref{tab:attention_configs} summarizes the attention-kernel
shapes, the shared sweep axes, and the CUTLASS GEMM shapes. The three
attention shapes are chosen to span the GQA-ratio axis.

\begin{table}[t]
\centering
\caption{Workload corpus. The $M$ of every GEMM shape
is swept over $\{16, 64, 256, 512, 2048, 8192\}$ in each of the three
datatypes, giving $18 \times 6 \times 3 = 324$ GEMM cells.}
\label{tab:attention_configs}
\footnotesize
\setlength{\tabcolsep}{3pt}
\renewcommand{\arraystretch}{1.15}
\begin{tabular*}{\linewidth}{@{}l@{\extracolsep{\fill}}ccc r@{}}
\toprule
\multicolumn{5}{@{}l}{\textbf{(a) Attention-kernel shapes}
 (head dim $D{=}D_v{=}128$, causal)} \\
\cmidrule(r){1-5}
\textbf{Model tag} & \textbf{Query} & \textbf{KV} & \textbf{GQA ratio}
 & \textbf{Production shape} \\
 & \textbf{heads} $H$ & \textbf{heads} $H_k$ & $H/H_k$ & \\
\midrule
\texttt{dense\_mha} & 32  & 32  & 1 & Llama-3-8B~\cite{llama3} \\
\texttt{gqa}        & 64  & 8   & 8 & Llama-3-70B~\cite{llama3} \\
\texttt{mla}        & 128 & 128 & 1 & DeepSeek-V2~\cite{deepseekv2} \\
\bottomrule
\end{tabular*}\\[3pt]
\begin{tabular*}{\linewidth}{@{}l@{\extracolsep{\fill}}l@{}}
\toprule
\multicolumn{2}{@{}l}{\textbf{(b) Attention sweep axes}} \\
\cmidrule(r){1-2}
Phase (query len \textsc{sq}) & decode (\textsc{sq}{=}1), chunked
 (\textsc{sq}{=}512), prefill (\textsc{sq}{=}\textsc{ctx}) \\
Context (KV len \textsc{ctx}) & 8K, 16K, 32K, 64K, 128K tokens \\
Page size & 16, 32, 64, 128, 256 tokens \\
Datatype & BF16, FP16, FP8 \\
\bottomrule
\end{tabular*}\\[3pt]
\begin{tabular*}{\linewidth}{@{}l@{\extracolsep{\fill}}r@{$\,\times\,$}l
 r@{$\,\times\,$}l r@{$\,\times\,$}l r@{$\,\times\,$}l@{}}
\toprule
\multicolumn{9}{@{}l}{\textbf{(c) CUTLASS GEMM, dense projections}
 ($N \times K$)} \\
\cmidrule(r){1-9}
\textbf{Family} & \multicolumn{2}{c}{\texttt{qkv\_proj}}
 & \multicolumn{2}{c}{\texttt{o\_proj}}
 & \multicolumn{2}{c}{\texttt{mlp\_gate}}
 & \multicolumn{2}{c}{\texttt{mlp\_down}} \\
\midrule
Llama-3-8B  &  6144 & 4096 & 4096 & 4096 & 14336 & 4096 & 4096 & 14336 \\
Llama-3-70B & 10240 & 8192 & 8192 & 8192 & 28672 & 8192 & 8192 & 28672 \\
DeepSeek-V2 &  7680 & 5120 & 5120 & 5120 & 12288 & 5120 & 5120 & 12288 \\
\bottomrule
\end{tabular*}\\[3pt]
\begin{tabular*}{\linewidth}{@{}l@{\extracolsep{\fill}}r@{$\,\times\,$}l
 r@{$\,\times\,$}l r@{$\,\times\,$}l@{}}
\toprule
\multicolumn{7}{@{}l}{\textbf{(d) CUTLASS GEMM, DeepSeek-V2 MLA
 projections} ($N \times K$)} \\
\cmidrule(r){1-7}
\textbf{Family} & \multicolumn{2}{c}{\texttt{q\_down}}
 & \multicolumn{2}{c}{\texttt{kv\_down}}
 & \multicolumn{2}{c}{\texttt{q\_up}} \\
\midrule
DeepSeek-V2 MLA & 1536 & 5120 & 512 & 5120 & 24576 & 1536 \\
\addlinespace[4pt]
 & \multicolumn{2}{c}{\texttt{k\_up}}
 & \multicolumn{2}{c}{\texttt{v\_up}}
 & \multicolumn{2}{c}{\texttt{o\_proj}} \\
\midrule
 & 16384 & 512 & 16384 & 512 & 5120 & 16384 \\
\bottomrule
\end{tabular*}
\end{table}

\subsection{Microbenchmark harness}

Each hypothesis in Section~\ref{sec:intro} names a single discriminating
feature (transfer size, rank, multicast degree, DSM reuse count,
SMEM pressure). For each, we build a self-contained CUDA kernel
that exercises only that feature and nulls out the others.
Conventions:

\textbf{Isolation.} Microbenchmarks for multicast scaling fix the
per-CTA tile shape and sweep cluster size 1--16 (H100), so that any
speedup is
attributable to multicast and not to a change in working-set
shape.

\textbf{Timing.} Wall-clock time uses CUDA event pairs over up to
20 warmup iterations followed by up to 200 timed iterations
(median reported, L2 flushed between iterations).

\textbf{Counters.} Per-mechanism architectural attribution uses
NCU~\cite{ncu_guide} counters. Counter collection
is run in a separate invocation from timing measurements.

\subsection{Sweep parameters}
\label{sec:method:sweeps}

\textbf{FA3 attention sweeps.} Three model shapes serve as proxies
for deployed configurations~\cite{llama3,deepseekv2}: dense MHA
for Llama-3-8B, GQA for Llama-3-70B, and MLA for DeepSeek-V2 (with
FlashMLA~\cite{flashmla} cited as the production MLA reference for
the page-block-size convention). For each model we sweep KV length
across a factor-of-two ladder spanning the long-context regime and
a short-context companion ladder, three execution phases (decode,
chunked prefill, full prefill), and a page-size grid chosen to
straddle the FA3 K-tile boundary for the datatypes BF16, FP16, and FP8. At
every TMA-eligible page size we run an equivalent comparison against a forced
\texttt{cp.async}-paged variant of the same kernel so the load
mechanism is the only varied factor.

\textbf{Microbenchmarks.} TMA 2D throughput sweeps SMEM tile
height and pipeline depth at a fixed tile width. TMA multicast
sweeps cluster size, sharing-group size, and SMEM tile shape, and
contrasts an explicit-multicast mode against a redundant-unicast 
mode at matched footprints to expose the multicast-vs-redundant-unicast
crossover. Finally, a DSM broadcast microbenchmark sweeps tile sizes of transfers within a cluster.

\section{Evaluation}
\label{sec:evaluation}

In this section, we dive into the evaluation of various optimization choices in the context of TMA on the workloads described in Section~\ref{sec:method}.

\subsection{Impact of KV page size}

\begin{figure}[t!]
  \centering
  \includegraphics[width=\columnwidth]{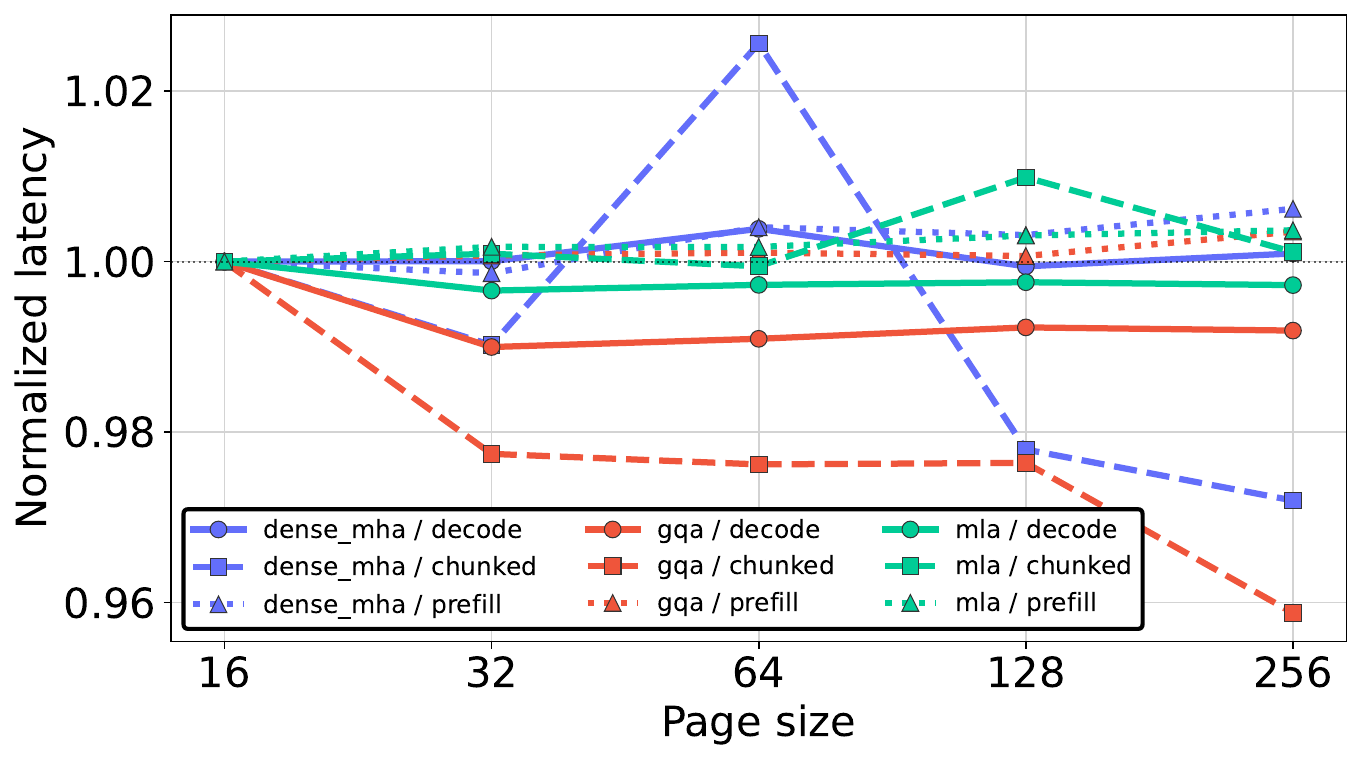}
  \caption{Median latency of the \texttt{cp.async}-paged FA3 forward
  at $\text{ctx}{=}131{,}072$, normalized per cell to the
  page-size-16 baseline. The nine curves span the model $\times$
  phase product (dense MHA, GQA, MLA) $\times$ (decode, chunked
  prefill, prefill) on H100 FP16.}
  \label{fig:page_size_sweep}
\end{figure}

Figure~\ref{fig:page_size_sweep} sweeps page size from 16 to 256 on
the \texttt{cp.async}-paged path. All nine $(\text{model},
\text{phase})$ curves stay within ${\pm}4\%$ of their page-size-16
baseline, with no systematic trend in page size. Over 50 timed iterations per point,
all 36 non-baseline cells have 95\% confidence intervals
within 2.5\% of their median. Page-size choice is therefore decoupled from kernel cost on either load path.

%The mechanism is structural: on
%the \texttt{cp.async} path each K-tile load is dominated by
%per-thread address computation and wavefront issue, while the
%page-table indirection itself is a single four-byte pointer load
%that hits L1 regardless of how many entries the table contains.

This decouples the page-size decision from kernel performance and
pushes it entirely to the allocator. The relevant tradeoff is
between tail-token waste at small page sizes and fragmentation
pressure at large ones, and it is workload-specific: long-context
workloads (RAG, code assistants, agent loops) keep their last page
mostly full, so the fragmentation cost of larger page sizes is
minor, and they should adopt page sizes at the TMA-dispatch
threshold to unlock the mechanism gain reported in
the next section. Short-context, high-concurrency workloads (chat,
function calling) keep many sequences whose last page is mostly
empty, and the per-sequence wasted tokens dominate the memory budget; for these, small page sizes
remain the right choice. 

%The rule of thumb is $\text{avg\_ctx} \geq 8 \times \text{page\_size}$.

\begin{observation}
\textbf{Page size is kernel-performance-neutral, but workload-specific.}
Forward-pass latency at fixed context is independent of page size on
both load paths to within noise. The choice is an allocator-level
tradeoff between fragmentation and TMA-dispatch eligibility, and it
should be made per workload: large pages for long-context serving,
small pages for short bursty traffic.
\end{observation}

\subsection{TMA vs. \texttt{cp.async} crossover point}
\label{sec:eval:crossover}

\begin{figure*}[t!]
\centering
\begin{minipage}[t]{0.32\linewidth}
  \centering
  \includegraphics[width=\linewidth]{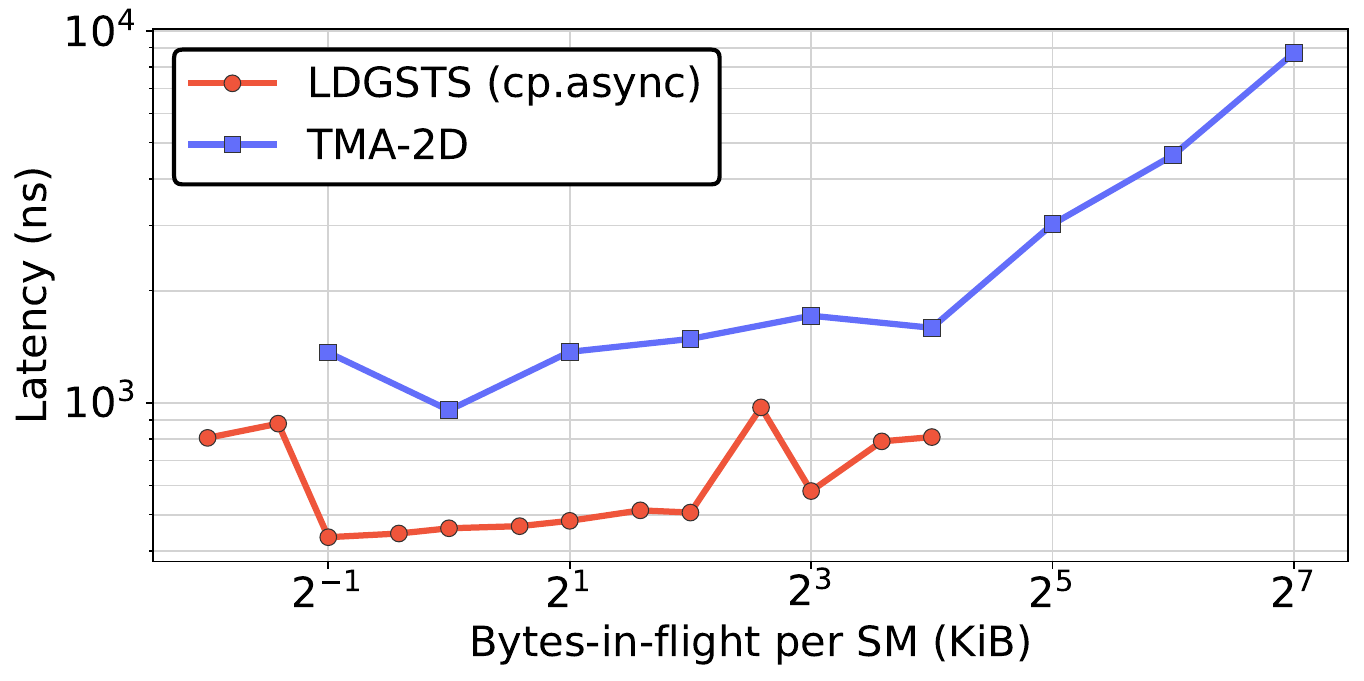}\\
  (a) Latency for data transfer size.
\end{minipage}\hfill
\begin{minipage}[t]{0.32\linewidth}
  \centering
  \includegraphics[width=\linewidth]{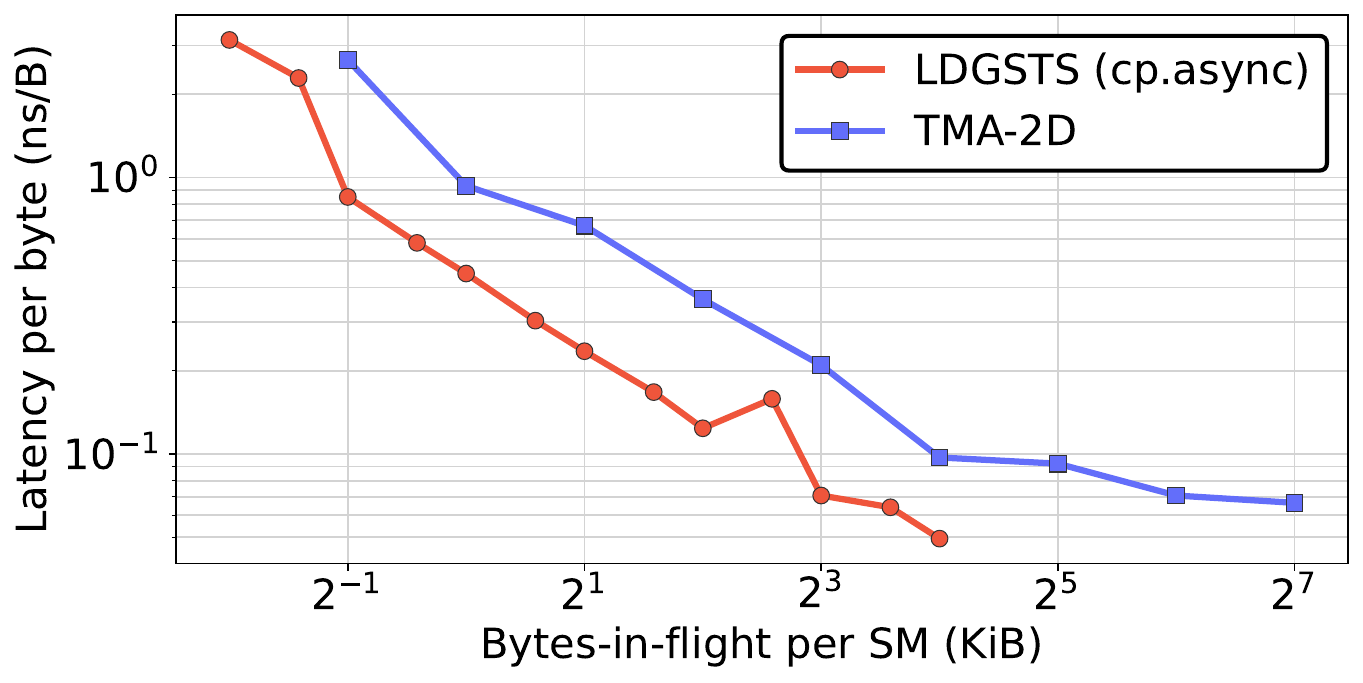}\\
  (b) Latency per byte (cost-amortization view).
\end{minipage}\hfill
\begin{minipage}[t]{0.32\linewidth}
  \centering
  \includegraphics[width=\linewidth]{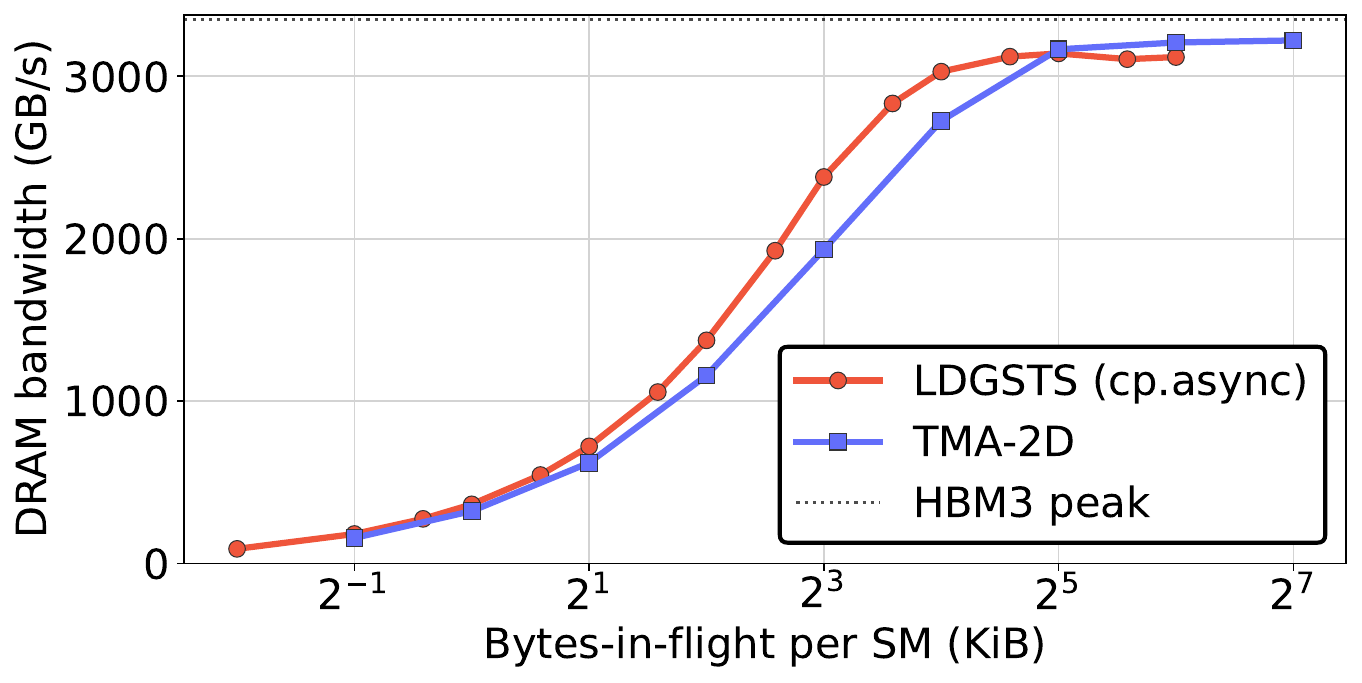}\\
  (c) Sustained throughput.
\end{minipage}
\caption{Latency and throughput microbenchmarks of LDGSTS
(\texttt{cp.async}) and TMA-2D on H100 SXM5, with both mechanisms
plotted against bytes-in-flight per SM. \emph{(a)} median round-trip
latency of a single load. \emph{(b)} Latency per byte, isolating fixed per-issue cost from
per-byte cost. \emph{(c)} max sustained DRAM bandwidth across the
config sweep; dotted line marks the H100 HBM3 advertised peak.}
\label{fig:ubench_crossover}
\end{figure*}

To pin down the mechanism behind the production-kernel results, we strip the comparison to its
microbench primitive: one LDGSTS issue per thread,
or one 2D TMA per CTA. The throughput companion sweeps the
saturation curve by pipelining many in-flight loads. Three findings
emerge, summarized in Figure~\ref{fig:ubench_crossover}.

\textbf{TMA carries a fixed per-issue tax.} In panel~(a), at the
smallest matched operating point of $0.5$\,KiB bytes-in-flight per
SM, TMA's median single-shot latency is $1.4$\,$\mu$s versus LDGSTS's
$0.43$\,$\mu$s, a $3.1{\times}$ tax. The gap narrows but does not
close as bytes-in-flight grow: at $1$\,KiB it is $2.1{\times}$, at
$2$--$8$\,KiB it stays in the $2.8$--$3.0{\times}$ range, and only at
$16$\,KiB does TMA come within $2{\times}$ of LDGSTS. The gap is the
descriptor walk, bulk-copy issue, and synchronization completion
machinery that TMA carries unconditionally. In panels~(a)
and~(b), LDGSTS latency jitters by ${\sim}10\%$ at the smallest
sizes, where the measurement sits near the noise floor; this
jitter is still far below the $2$--$3{\times}$ latency gap TMA
carries in the same range.

\textbf{The fixed tax amortizes hard.} Panel~(b) shows ns/byte
collapsing by $\sim$40$\times$ over the swept range for \emph{both}
mechanisms, with LDGSTS holding a $2$--$3{\times}$ per-byte advantage
across the $0.5$--$16$\,KiB range. As bytes-in-flight grow, both
curves converge toward the same per-byte floor: TMA reaches
$\sim$0.07\,ns/B at $128$\,KiB, statistically indistinguishable from
LDGSTS's $\sim$0.05\,ns/B at its $16$\,KiB ceiling. The descriptor
cost is genuinely a fixed overhead, not a per-byte one, so it
disappears as soon as the kernel keeps enough bytes in flight per
issue.

\textbf{The HBM3 ceiling is reached by both, but TMA crosses it
without pipelining.} Panel~(c) shows that LDGSTS gets to $\sim$3.1~TB/s
(94\% of HBM3 peak) at 32~KiB Bytes in Flight (BIF), but only by stacking
4~CTAs/SM\(\times\)stages\(\times\)256~threads\(\times\)float4 to
fabricate that many in-flight bytes. TMA reaches a slightly higher
$\sim$3.2~TB/s (96\%) at 32~KiB BIF using a single
CTA's single descriptor with a few pipeline stages. \emph{The
crossover point at which TMA becomes the right choice is therefore
not a throughput point but an architectural one}: when the per-CTA
work cannot mask the LDGSTS issue stream, or when a
single tile exceeds what 256 threads can address with one float4
LDGSTS each ($\sim$16\,KiB), TMA is the only path that delivers the bandwidth
in one issue.

\begin{observation}
\textbf{TMA wins only past a per-issue payload threshold.} TMA's
latency is $2$--$3{\times}$ LDGSTS's at the smallest
payloads where both mechanisms are measurable ($0.5$--$8$\,KiB
bytes-in-flight per SM), converges within $2{\times}$ at $16$\,KiB,
and reaches statistical parity in per-byte cost at $\geq$$32$\,KiB.
Two distinct crossover thresholds matter for kernel authors: the
\emph{per-byte cost} curves meet near $16$\,KiB, and the
\emph{sustained throughput} curves saturate the same HBM3 ceiling at
$\sim$$32$\,KiB. Below $16$\,KiB bytes-in-flight per SM,
\texttt{cp.async} is faster end-to-end; above $32$\,KiB, TMA
saturates the same DRAM ceiling at a fraction of the issue-side
machinery, leaving more warps free for compute.
\end{observation}

\subsection{TMA vs. \texttt{cp.async} in production kernels}
\label{sec:eval:tma-vs-cpasync}

\begin{figure*}[t!]
\centering
\includegraphics[width=175mm]{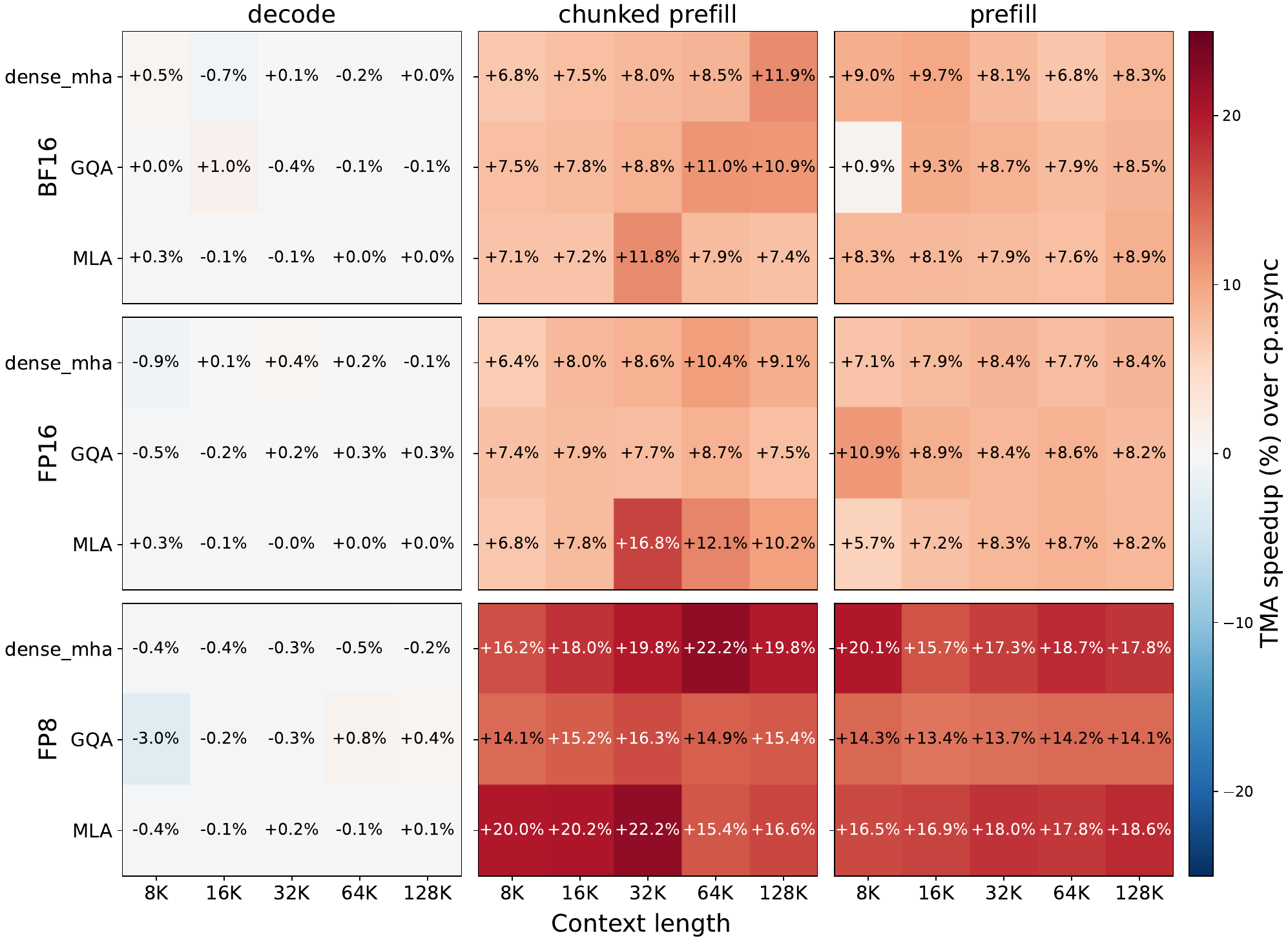}
\caption{TMA-paged speedup over forced \texttt{cp.async}-paged on
  the FA3 forward at the smallest TMA-eligible page size per datatype
  (BF16 and FP16: $\text{ps}{=}128$; FP8: $\text{ps}{=}192$). Each cell reports
$1-t_{\text{TMA}}/t_{\text{cp.async}}$ as a percentage on H100,
causal, across the three model shapes (dense MHA, GQA, MLA), the
three execution phases (decode, chunked prefill, prefill), and the
KV-length ladder
$\{8\text{K},16\text{K},32\text{K},64\text{K},128\text{K}\}$.}
\label{fig:tma_cp_heatmap}
\end{figure*}

Figure~\ref{fig:tma_cp_heatmap} sweeps the TMA-vs-\texttt{cp.async}
speedup over the model, phase, datatype, and KV-length axes at the
TMA-eligible page size. The pattern is sharply phase-dependent.
\emph{Decode} cells sit on the zero line: the kernel is HBM-bandwidth-bound and both load
paths move the same bytes from DRAM, so the TMA descriptor-and-tile
machinery has nothing to amortize. \emph{Chunked prefill} and
\emph{prefill} cells, by contrast, are compute-bound and TMA
delivers a 5.7--11.9\% speedup in BF16/FP16 and a
much larger 13.4--22.2\% in FP8. At 95\% confidence
the prefill mean speedups are $+7.9\%$ $[+6.9, +8.9]$ (BF16),
$+8.2\%$ $[+7.6, +8.8]$ (FP16), and $+16.5\%$ $[+15.5, +17.5]$
(FP8), and every prefill cell is individually positive; decode ties
zero, with intervals of $[-0.18\%, +0.22\%]$ (BF16),
$[-0.17\%, +0.16\%]$ (FP16), and $[-0.72\%, +0.12\%]$ (FP8).
The bump in speedup due to TMA stems from a more efficient use of warp issue bandwidth. TMA keeps memory accesses in flight using fewer issue slots, freeing the rest for compute. The extra FP8 headroom traces to FA3's tile
scheduler: at FP8/hd=128 it picks a larger tile size (192) in the
prefill mainloop (vs.\ 128 in BF16/FP16), so each TMA issue moves a
50\% larger tile and the per-issue overhead is amortized harder,
while the smaller bytes-per-token of FP8 lets compute keep up at
the wider tile.

\textbf{Under-the-hood: instruction-level mechanism.}
Table~\ref{tab:ncu_triplet} contrasts NCU counters for TMA-paged
against \texttt{cp.async}-paged on three canonical FA3 operating
points spanning the compute-to-memory-bound spectrum:
\emph{TC-bound} at dense\_mha/prefill/FP16/ctx=16\,K (79\% tensor
pipe, 5.7\% of HBM peak), \emph{Balanced} at
MLA/chunked/FP8/ctx=7680 (compute and memory both with headroom),
and \emph{HBM-bound} at dense\_mha/decode/FP16/ctx=128\,K (92\% of
HBM3 peak). Three observations tie back to the
Fig~\ref{fig:tma_cp_heatmap} pattern.
\emph{First}, TMA and \texttt{cp.async} move identical DRAM bytes and
issue identical WGMMA counts in every regime, so the mechanism
difference is not about data movement but about the instruction
stream that surrounds it: TMA replaces per-thread LDGSTS with bulk
TMA-pipe issues at a consistent 26--33$\times$ compression across all
three regimes, and hoists per-thread address arithmetic out of the
mainloop (FMA and ALU pipes drop 8--22\%).
\emph{Second}, wall-time payoff tracks which pipe is on the critical
path: at TC-bound and Balanced, the freed warp-scheduler bandwidth
cashes out as +7.8~pp higher tensor-pipe utilization in both and
10--21\% shorter GPC cycles. At HBM-bound decode, DRAM is at 92\% of
HBM3 peak in both arms, so mechanism choice cannot buy any wall time
even though the same 15$\times$+ instruction-count compression is
present.
\emph{Third}, this is why FA3's runtime dispatch heuristic disables
TMA-paged at decode by default: the instruction-side compression is
still delivered when TMA is dispatched, but it produces no wall-time
change when DRAM is on the critical path (HBM-bound column).

\subsubsection{Generalization to CUTLASS GEMM}
\label{sec:eval:cutlass}

To test whether the mechanism generalizes beyond attention, we repeat
the TMA-vs-\texttt{cp.async} comparison on 324 CUTLASS SM90 GEMM cells (18
shapes $\times$ 6 $M$-values $\times$ 3 datatypes, listed in
Table~\ref{tab:attention_configs}). TMA wins in 319 of 324 cells at a
median 1.24$\times$ speedup, driven by the same memory-issue
compression seen in the FA3 counter table (30--48$\times$ across GEMM
regimes). The largest wins (up to 1.31$\times$) occur where the
\texttt{cp.async} arm is SM-utilization-limited by its issue stream;
wins shrink toward 1.00$\times$ as GEMMs approach the HBM ceiling,
matching the FA3 decode ceiling. The 5 losing cells are all at the
smallest $M$ ($M \in \{16, 64\}$) where too few bytes-in-flight per SM
prevent the descriptor overhead from amortizing, and TMA is 2--3\%
slower---the same crossover threshold observed in the
Fig~\ref{fig:ubench_crossover} microbenchmarks. The CUTLASS result
confirms that the TMA dispatch rules derived from the FA3 catalog
generalize to the operand-broadcast GEMM workload without
qualitative change.

\begin{observation}
\textbf{The TMA advantage is a single mechanism with regime-dependent payoff.}
Across FA3 attention and CUTLASS GEMM, TMA replaces per-thread memory
issues with bulk hardware issues at an order-of-magnitude compression
in every regime tested; what varies is whether that compression
converts to wall time savings, which depends on the critical
path. At decode, attention is HBM-bandwidth-bound and TMA ties
\texttt{cp.async} to within noise. At chunked prefill and prefill,
which are compute-bound, TMA delivers a 5.7--11.9\% wall-clock win
in BF16/FP16 and 13.4--22.2\% in FP8. The win grows with the
per-tile work the datatype lets the
scheduler pack: FP8 gains extra ground because FA3 selects a 50\%
wider tile, amortizing the descriptor and per-issue overhead over
more bytes per load. Kernels that straddle the crossover (small tiles
at small $M$) can instead pay a small penalty when the descriptor
cost fails to amortize.
\end{observation}

\begin{table*}[t!]
\centering
\caption{Per-kernel NCU counters comparing TMA against
\texttt{cp.async} on three representative FA3 forward regimes.
Bolded rows identify the mechanism-distinguishing counters and wall
time.}
\label{tab:ncu_triplet}
\small
\renewcommand{\arraystretch}{1.15}
\begin{tabular}{l | r r c | r r c | r r c}
\hline
& \multicolumn{3}{c|}{\textbf{TC-bound} (prefill/16K/FP16)}
& \multicolumn{3}{c|}{\textbf{Balanced} (chunked/8K/FP8)}
& \multicolumn{3}{c}{\textbf{HBM-bound} (decode/128K/FP16)} \\
\textbf{Metric} & \textbf{TMA} & \texttt{cp.async} & \textbf{ratio}
                & \textbf{TMA} & \texttt{cp.async} & \textbf{ratio}
                & \textbf{TMA} & \texttt{cp.async} & \textbf{ratio} \\
\hline
Tensor pipe (\% peak)     & \textbf{79.3\%} & 71.5\% & \textbf{1.11$\times$} & 37.9\% & 30.1\% & \textbf{1.26$\times$} & 25.4\% & 25.5\% & 1.00$\times$ \\
SM active (\% peak)       & 99.5\%  & 99.5\%  & 1.00$\times$ & 95.1\%  & 95.5\%  & 1.00$\times$ & 91.0\%  & 91.3\%  & 1.00$\times$ \\
DRAM bytes                & 689\,M  & 690\,M  & 1.00$\times$ & 327\,M  & 327\,M  & 1.00$\times$ & 2.15\,G & 2.15\,G & 1.00$\times$ \\
DRAM throughput (\%)      & 5.7\%   & 5.2\%   & 1.10$\times$ & 22.5\%  & 18.0\%  & 1.25$\times$ & \textbf{92.1\%} & \textbf{92.4\%} & 1.00$\times$ \\
L2 hit rate (\%)          & 94.2\%  & 94.0\%  & 1.00$\times$ & 69.2\%  & 73.5\%  & 0.94$\times$ & 1.5\%   & 1.5\%   & 1.00$\times$ \\
\textbf{Total inst issued}         & \textbf{1.09\,G} & \textbf{1.26\,G} & \textbf{0.86$\times$} & \textbf{145.6\,M} & \textbf{166.0\,M} & \textbf{0.88$\times$} & \textbf{66.5\,M} & \textbf{87.4\,M} & \textbf{0.76$\times$} \\
\quad WGMMA               & 33.8\,M & 33.8\,M & 1.00$\times$ & 1.6\,M  & 1.6\,M  & 1.00$\times$ & 2.1\,M  & 2.1\,M  & 1.00$\times$ \\
\quad \textbf{Memory} & \textbf{1.3\,M} & \textbf{34.1\,M} & \textbf{0.04$\times$} & \textbf{57\,K}  & \textbf{1.9\,M}  & \textbf{0.03$\times$} & \textbf{131\,K} & \textbf{4.2\,M}  & \textbf{0.03$\times$} \\
\quad FMA pipe            & 451\,M  & 506\,M  & 0.89$\times$ & 66\,M   & 72\,M   & 0.92$\times$ & 29\,M   & 35\,M   & 0.83$\times$ \\
\quad ALU pipe            & 260\,M  & 291\,M  & 0.89$\times$ & 36\,M   & 39\,M   & 0.91$\times$ & 16\,M   & 21\,M   & 0.78$\times$ \\
\hline
\textbf{GPC cycles (wall time)}    & \textbf{5.17\,M} & \textbf{5.73\,M} & \textbf{0.90$\times$} & \textbf{617\,K}  & \textbf{778\,K}  & \textbf{0.79$\times$} & \textbf{993\,K}  & \textbf{993\,K}  & \textbf{1.00$\times$} \\
\hline
\end{tabular}
\end{table*}

\subsection{TMA multicast vs redundant unicast}
\label{sec:eval:tma-multicast}

\begin{figure*}[t!]
\centering
\begin{minipage}[t]{0.49\linewidth}
  \centering
  \includegraphics[width=\linewidth]{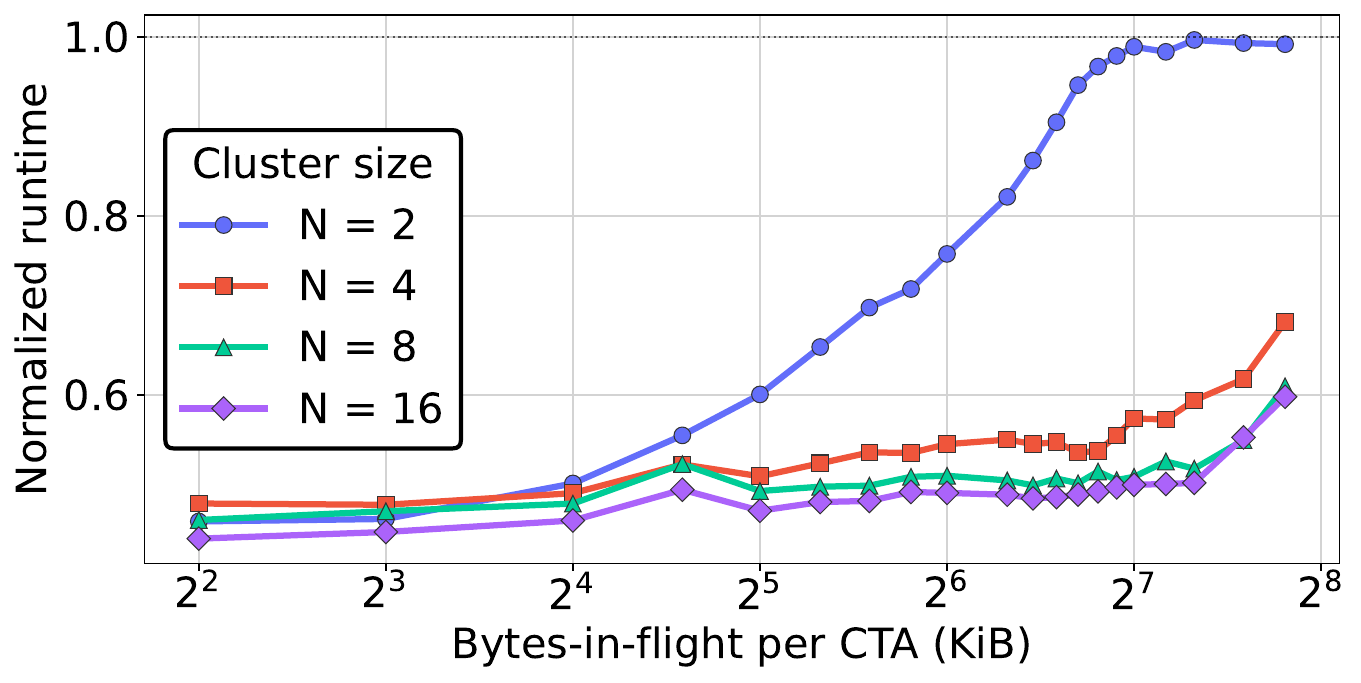}
  \\(a) Kernel runtime for redundant unicast normalized to TMA multicast (< 1 means unicast is faster).
\end{minipage}\hfill
\begin{minipage}[t]{0.49\linewidth}
  \centering
  \includegraphics[width=\linewidth]{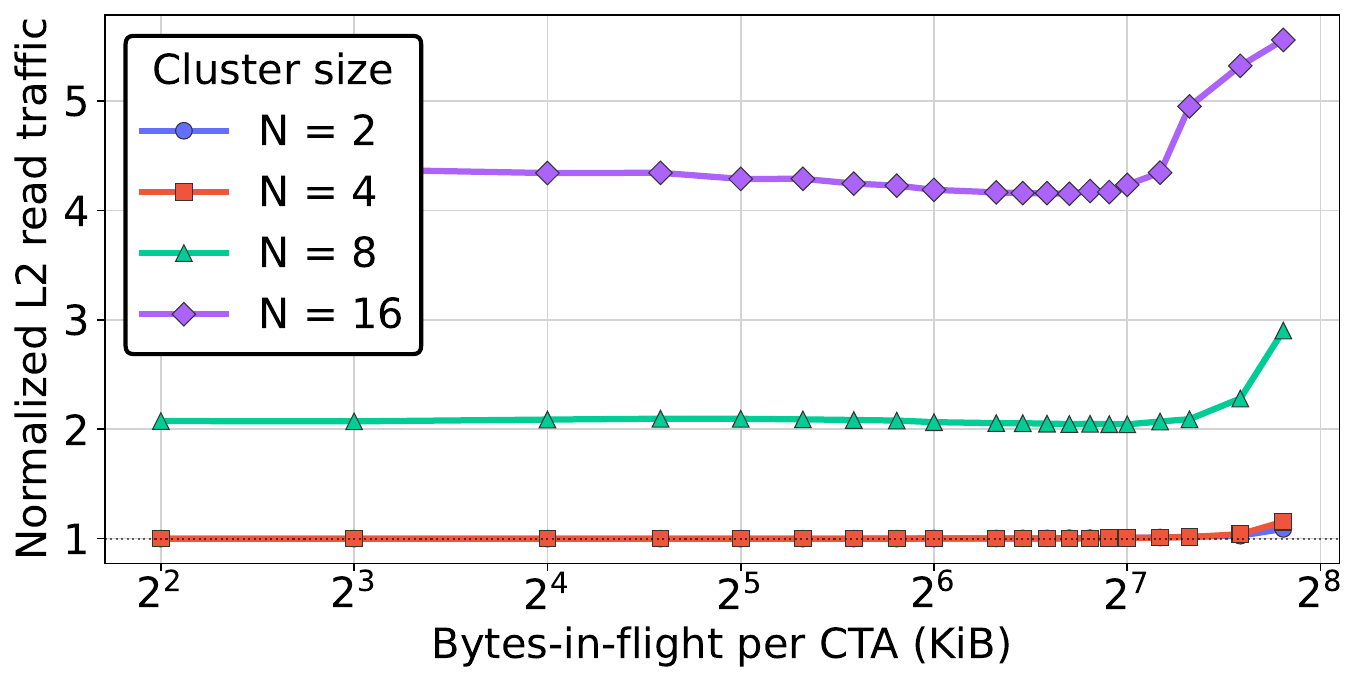}
  \\(b) L2 read sectors served for redundant unicast normalized to TMA multicast (> 1 means unicast suffers more L2 traffic).
\end{minipage}
\caption{Normalized runtime and L2 reads for redundant unicast vs.\ explicit TMA multicast on H100. Each cluster of $N$ CTAs co-located in a thread-block cluster
reads the same tile per iteration; the redundant unicast arm has every CTA issue its
own unicast TMA and lets the GPC L2 Request Coalescer (LRC) merge
duplicates; the explicit arm uses
TMA multicast from the leader CTA.}
\label{fig:tma_multicast}
\end{figure*}

When the same tile is read by multiple CTAs, Hopper offers two paths to
collapse the duplication: \emph{(i)} let each CTA issue its own unicast TMA
and rely on the GPC's LRC to deduplicate the reads (redundant unicast), or \emph{(ii)} have the leader CTA issue a single
multicast TMA whose hardware fan-out delivers the tile into all peer SMEMs
(TMA multicast). The two mechanisms move identical bytes from DRAM, but
they expose very different upstream traffic and very different
synchronization requirements.

Figure~\ref{fig:tma_multicast} sweeps both arms across tile size and cluster
size~$N$, normalizing the redundant unicast arm to the TMA multicast. Panel~(a) shows the normalized kernel runtime: redundant unicast is
$\mathbf{0.45\!-\!0.79{\times}}$ the multicast latency at every $N\ge 2$, a
$1.3\!-\!2.2{\times}$ speedup that grows with $N$ and saturates near
$2{\times}$ at $N\ge 8$. The win is not the load itself but its absence of
coordination: the multicast protocol requires a cluster-level synchronization
which sits on the critical path. Redundant unicast
uses per-CTA-local mbarriers and runs free.

Panel~(b) shows the normalized L2 read-sector traffic. At $N\le 4$ the
two arms tie within 1\%: the LRC's coalescing window comfortably absorbs the
inter-CTA arrival skew, and L2 sees a single sector per tile regardless of
which path issued the loads. At $N=8$ redundant unicast leaks
$2.1{\times}$ more L2 sectors; at $N=16$, $\mathbf{4.4{\times}}$ more.
The skew exceeds the LRC's tracking depth, misses deduplication opportunities, and the request
bandwidth at L2 scales linearly with $N$ on the redundant unicast path while
TMA multicast holds a constant 1 sector per tile per iteration by construction. Table
\ref{tab:multicast_decision} summarizes the decision rule for choosing between explicit TMA multicast and redundant unicast on Hopper.

\begin{table}[t]
\centering
\caption{Workload-specific decision rule for choosing between
explicit TMA multicast and redundant unicast on Hopper.}
\label{tab:multicast_decision}
\renewcommand{\arraystretch}{1.25}
\begin{tabular}{p{0.42\linewidth} p{0.48\linewidth}}
\hline
\textbf{Condition} & \textbf{Recommended primitive} \\
\hline
Cluster $N=2$ AND bytes-in-flight per CTA $\gtrsim 100$~KiB
(largest tiles in the sweep, Fig.~\ref{fig:tma_multicast}(a))
 & Either path; ratio approaches~$1.0$ and L2 traffic is tied.
 The extra implementation complexity of multicast is not
 justified. \\
Cluster $N\le 4$ at smaller tiles, OR any $N\ge 8$ where the
kernel is streaming / not L2-bound 
 & \textbf{Redundant unicast} Redundant unicast
 is $1.3{-}2.1{\times}$ faster on runtime because the cluster-level synchronization required by the TMA multicast protocol
 dominates the critical path; the extra L2 sectors that LRC
 leaks at $N\ge 8$ are absorbed without cost when L2 is just a
 pass-through to DRAM. \\
Cluster $N\ge 8$ AND kernel is L2-bandwidth-bound or reuses
tiles across passes (\textit{e.g.}, CUTLASS GEMM operand
broadcast)
 & \textbf{Explicit TMA multicast.} The $2{-}4{\times}$ L2
 sector savings (Fig.~\ref{fig:tma_multicast}(b)) translate to
 wall-time gain once L2 is on the critical path, outweighing
 the sync overhead. \\
\hline
\end{tabular}
\end{table}

\begin{observation}
\textbf{TMA multicast trades runtime for L2 cleanliness.} Letting the LRC deduplicate
N CTAs' independent TMA issues is up to $2{\times}$ faster end-to-end than
explicit multicast because the cluster sync that multicast requires
dominates the critical path. The cost is L2 request bandwidth: at
cluster~$\ge 8$, implicit dedup spends $2{-}4{\times}$ more L2 sectors per
tile because the inter-CTA skew breaks LRC coalescing. Choose explicit
multicast when the surrounding kernel is L2-bandwidth-bound or reuses
tiles across passes (\textit{e.g.}, CUTLASS GEMM operand broadcast); choose
redundant unicast when the workload is streaming and L2 is a pass-through to
DRAM.
\end{observation}

\subsection{TMA multicast vs DSM-based broadcast}
\label{sec:eval:tma-multicast-vs-dsm}

Hopper exposes two hardware primitives that solve the same logical
problem: get one GMEM tile resident in the SMEM of $N$ CTAs of a
thread-block cluster. \emph{TMA multicast} accomplishes this with a
single \texttt{cp.async.bulk.tensor.\-multicast::cluster} issue from a
leader CTA; the TMA engine pulls the tile once from L2 and uses the
SM-to-SM crossbar to fan it out, pushing it into each peer's SMEM and
signaling each peer's \texttt{mbarrier}. \emph{DSM-based broadcast} accomplishes the same end state with two
software steps: the leader issues a unicast TMA that lands the tile in
its own SMEM, after which peer CTAs read the tile through the same
SM-to-SM crossbar via \texttt{ld.shared::cluster} instructions and potentially
stage it into their own SMEM. Both protocols move the same bytes from
DRAM to N peer SMEMs and pay one DRAM trip per tile; they differ only
in how the cross-cluster step is executed: \emph{push}, hardware
driven, single issue (multicast); or \emph{pull}, software driven, $N{-}1$
parallel issues bracketed by a cluster-level synchronization primitive on each side
(DSM).

Multicast issues a single
multicast TMA from the leader. DSM issues a unicast TMA on the
leader, performs a cluster-wide barrier, has peers pull the tile via
cluster-scope shared-memory loads and stage it into their local SMEM. We sweep cluster
size $N \in \{2, 4, 8, 16\}$ and 20 tile sizes from 1~KiB to 56~KiB. Both protocols
incur the same DRAM cost per tile, so any wall-time delta is purely
the cross-cluster fan-out mechanism.

\begin{figure}[t]
\centering
\includegraphics[width=\columnwidth]{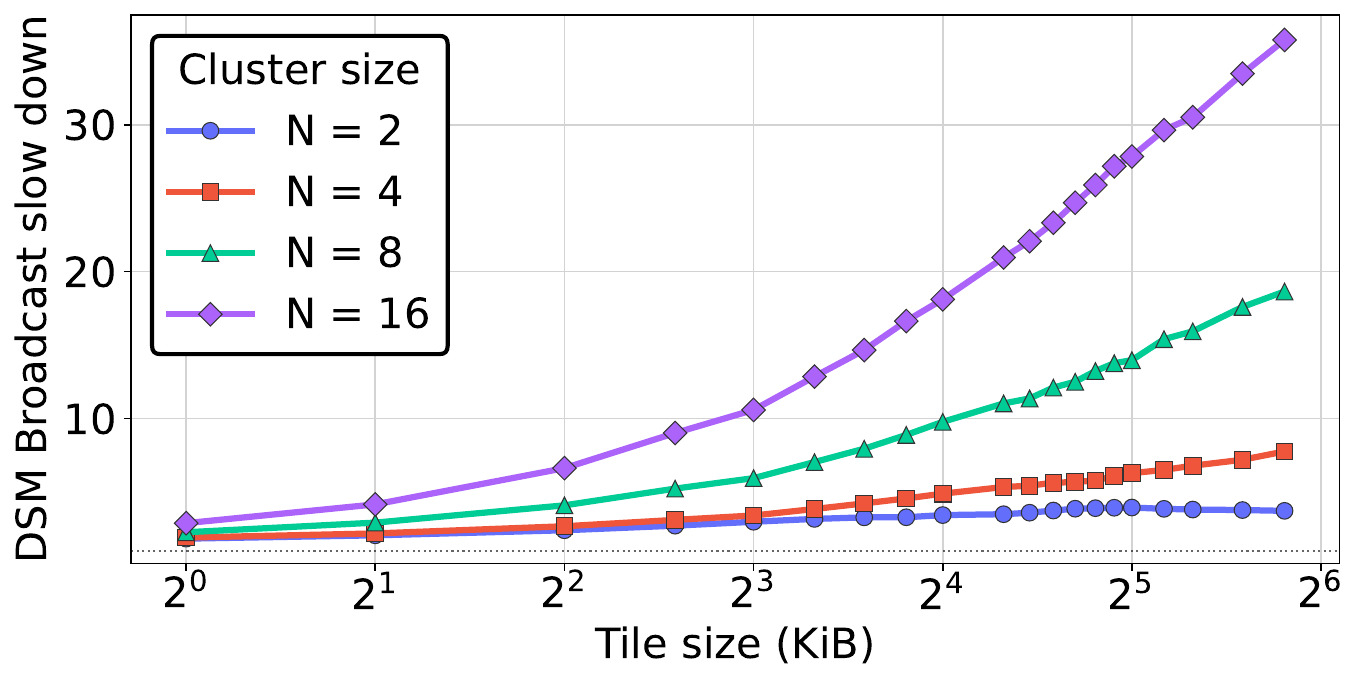}
\caption{Normalized slowdown of DSM broadcast with respect to TMA
multicast at various cluster sizes and tile sizes.
Ratio $>1$ means TMA multicast is
faster.}
\label{fig:dsm_vs_mcast}
\end{figure}

Figure~\ref{fig:dsm_vs_mcast} shows the slowdown of DSM broadcast with respect to TMA multicast. TMA multicast is
uniformly faster at every cell, and the gap widens monotonically along
both axes. At $N{=}2$ the wall-time ratio sits in the
$1.8{-}3.9{\times}$ range and is essentially flat in tile size; at
$N{=}4$ it grows from $1.9{\times}$ at the smallest tile to
$7.8{\times}$ at the largest; at $N{=}8$ from $2.3$ to
$\mathbf{18.6{\times}}$; and at $N{=}16$ from $2.9$ to
$\mathbf{35.8{\times}}$. Two scaling regimes explain the shape.
\emph{First}, the small-tile floor of $\sim$$2{-}3{\times}$ across
all $N$ is the structural overhead of the DSM protocol's higher synchronization overhead. At small tiles, where each
iteration's data movement is short, this synchronization tax dominates.
\emph{Second}, the large-tile divergence reflects the asymmetric
crossbar cost: DSM moves $(N{-}1)\times\mathrm{tile\_bytes}$ over the
SM-to-SM crossbar via $(N{-}1)$ independent software-driven pull
streams, while multicast issues a single hardware-driven fan-out from
a single L2 fetch. DSM's cross-cluster traffic scales linearly with
$N{-}1$ and with tile size; multicast's amortizes both factors into
one issue. The $\sim$36$\times$ corner at $(N{=}16, 56\,\text{KiB})$
is the consequence of both effects multiplying.

\begin{observation}
\textbf{TMA multicast dominates DSM broadcast on Hopper when the
broadcast pattern is regular.} TMA multicast is
$1.8{\times}$ to $35{\times}$ faster than a DSM-based broadcast across
all measured cluster sizes and tile sizes, with the gap growing in
both. The mechanism gap has two components: a fixed
$\sim$$2{-}3{\times}$ synchronization-overhead floor, and a
linear-in-$(N{-}1)$ crossbar-traffic penalty that dominates at large
tiles. DSM remains the right primitive when (i) the data is already
SMEM-resident (no DRAM trip to amortize), (ii) the consumer set is
irregular and not addressable as a single multicast mask, or (iii)
SMEM capacity dictates a single shared copy rather than $N$ replicas.
For the regular broadcast pattern that CUTLASS and FA3 actually use
(operand tile or KV tile to every cluster CTA), TMA multicast is the
correct hardware choice on Hopper.
\end{observation}

\section{Extension to Blackwell (SM100)}
\label{sec:blackwell}

To test whether the H100 dispatch rules generalize, we repeat the
study on CUTLASS GEMMs (Sec.~\ref{sec:eval:cutlass}
Table~\ref{tab:attention_configs}) on the Blackwell (SM100) B200. B200 retains the TMA, cluster, multicast, and DSM mechanisms
of Hopper but adds Tensor Memory (TMEM), which moves the UMMA
accumulator state out of SMEM (freeing $\sim$256\,KB per SM for larger
TMA tiles), and a 2-CTA UMMA that changes per-issue tile requirements.

\textbf{TMA wins widen on B200.} TMA wins 320 of 324 cells at a median
$2.12\times$, up from $1.24\times$ on H100
(Sec.~\ref{sec:eval:cutlass}); the 4 ties sit in the small-$M$
crossover and no cell shows a loss. Two SM100 shifts explain the
widened advantage: (i) per-SM tensor throughput roughly doubles while
LDGSTS retires through the unchanged per-thread issue machinery, so
\texttt{cp.async} can no longer feed the tensor cores and the TMA
arm's issue-side compression (Table~\ref{tab:ncu_triplet}) cashes out
at a larger ratio; and (ii) the HBM3e bandwidth of
B200~\cite{nvidia_blackwell} outgrows kernel demand so no cell
saturates HBM, emptying the DRAM-bound tie regime that existed on H100.

The reported gain is a lower bound: we use the 1-SM UMMA path to keep
both arms comparable, since 2-SM UMMA is TMA-only. The qualitative
dispatch rules (payload threshold, compute-bound gate, multicast
preconditions) transfer to B200 without change; only the absolute
crossovers shift, tracking tensor-core and bandwidth scaling rather
than copy-engine improvements. Extending the analysis to production
FlashAttention-4~\cite{flashattention4} on B200 remains future work.

\section{Related Work}
\label{sec:related}

Our characterization touches six prior threads; none resolves the
comparative dispatch question we frame.

\textbf{Memory-hierarchy microbenchmarking of Hopper and Blackwell.}
Luo et al.~\cite{luo_hopper,luo_hopper_arxiv} characterize the TMA,
L1TEX, L2, and DSM paths on H100 in isolation, and Jarmusch et
al.~\cite{jarmusch_blackwell} extend that style of dissection to
Blackwell; L\"{u}hnen et al.~\cite{luhnen_thesis} report
complementary measurements of the cluster crossbar.
These works bound each mechanism alone, but never contrast TMA
against \texttt{cp.async} on the same operand or multicast against
DSM on the same broadcast workload.

\textbf{Asynchronous memory copy on Ampere and Hopper.} Li et
al.~\cite{li_asyncmemcpy,li_taco_pipelining} characterize the
\texttt{cp.async} (LDGSTS) path on A100 and H100, mapping its
overlap envelope and per-instruction issue cost. Their measurements
bound one side of our TMA-vs-\texttt{cp.async} comparison but never
contrast the two paths directly on identical operands.

\textbf{Production kernels and libraries using TMA.} Production
libraries encode TMA dispatch heuristics without publishing the
empirical crossovers behind them. CUTLASS~\cite{cutlass} uses TMA
and multicast for GEMM operand tiles;
FlashAttention-3~\cite{flashattention3} routes Q, K, and V through
TMA on prefill but gates TMA-paged-KV behind a workload-dependent
fallback to \texttt{cp.async}, FlashAttention-4~\cite{flashattention4}
revisits the same dispatch on SM100, and
FlashInfer~\cite{flashinfer}, FlashMLA~\cite{flashmla}, and
cuBLAS~\cite{cublas} make analogous operand-by-operand
TMA-vs-fallback choices. The FA3 in-source
dispatch comment~\cite{flashattention3_repo} and the
Colfax~\cite{colfax_tma_tutorial} and
PyTorch~\cite{pytorch_tma_fp8} write-ups all report TMA effects
without isolating the mechanism.

\textbf{KV cache memory management.}
PagedAttention~\cite{pagedattention}, vLLM~\cite{vllm}, SGLang~\cite{sglang},
TensorRT-LLM~\cite{trtllm}, and vAttention~\cite{vattention} treat
page size as an allocator-level tradeoff and do not model its
interaction with TMA-eligibility on FA3 and FA4. Operator-default
page sizes fall back to \texttt{cp.async}-paged on H100, leaving
measurable TMA bandwidth unused.

\textbf{Vendor documentation and community reports.}
NVIDIA's Hopper Tuning Guide~\cite{nvidia_hopper_tuning} describes
the API but not crossover points. Developer-Forum
reports~\cite{forum_multicast_slow,forum_tma_oddity} document
deployed anomalies that our cross-cutting case study explains via
TMA-engine occupancy and leader-side issue serialization.

\textbf{Prior Blackwell characterization.}
Jarmusch et al.~\cite{jarmusch_blackwell} dissect the memory hierarchy
and SM pipelines of consumer Blackwell, and Chen and
Patel~\cite{semianalysis_blackwell_tensor} provide per-mechanism
baseline measurements for server Blackwell (B200,
SM100)~\cite{nvidia_blackwell}, bounding TMA, TMEM, and the 2-CTA
UMMA path in isolation. Section~\ref{sec:blackwell} builds on these
baselines to extend our workload-grounded characterization from H100
to B200.

To our knowledge, no prior work pairs a workload-grounded view of
TMA dispatch choices in a production attention kernel with
microbenchmarks that explain why those choices were made. Our paper distills the resulting
crossovers into concrete rules of thumb for TMA and multicast
dispatch on the Hopper generation of NVIDIA GPUs.

\section{Conclusion}
\label{sec:conclusion}

TMA is used pervasively in production Hopper kernels, yet its
deployment is conspicuously selective and the heuristics that gate
dispatch are workload-specific and under-documented. We closed that
gap with a workload-grounded study in three stages: an
operand-by-operand catalog of the TMA, multicast, DSM, and
\texttt{cp.async} choices that FlashAttention-3 makes; targeted
microbenchmarks on H100 that isolate the discriminating axes; and
a cross-cutting paged-KV case study that exercises every rule on a
deployed kernel.

The study answers the three questions we set out to address.
\emph{(Q1)} TMA outperforms \texttt{cp.async} only above a joint
crossover in transfer size and descriptor reuse; outside
that envelope (small transfers, low reuse), \texttt{cp.async} is the correct choice. The
FA3 sweep comparing TMA-paged against \texttt{cp.async}-paged across
model shapes, datatypes, and phases confirms that within either dispatch the
page-size knob is performance-neutral within measurement noise;
mechanism, not page size, is the load-bearing axis, and operators
of vLLM, SGLang, and TensorRT-LLM can choose page size on
allocator grounds and TMA dispatch eligibility (which disproportionately benefits compute-bound phases such as prefill operations). \emph{(Q2)} TMA multicast beats the
LRC-deduplicated redundant-unicast path when the kernel is L2 bandwidth bound; outside that regime,
redundant unicast is competitive end-to-end and multicast's L2
cleanliness is not load-bearing. \emph{(Q3)} Against DSM broadcast
at matched DRAM cost, multicast wins under high per-tile reuse,
while DSM wins under SMEM pressure, low reuse, or under exotic data transfer patterns.

Section~\ref{sec:blackwell} extends the CUTLASS GEMM sweep to server
Blackwell (B200, SM100), where TMA wins 320 of 324 cells at median
$2.12\times$ (up from $1.24\times$ on H100). The
qualitative dispatch rules carry across generations; the absolute
crossovers shift with tensor-core throughput and HBM bandwidth
scaling rather than with copy-engine changes per se. As future work,
we plan to extend the analysis to production attention kernels such as
FlashAttention-4~\cite{flashattention4} on B200. More broadly, we
hope that deriving dispatch rules from a workload-grounded
characterization, rather than from per-mechanism
peak numbers alone, encourages similar studies of the asynchronous
data-movement primitives that each new GPU generation introduces.

\section*{Acknowledgments}

This work was supported, in part, by NSF CCF \#1943379 and NSF CCF
\#2450447. We used Claude to write some of the scripts that drive our
experiments and to polish the text. All experimental results were produced by
running the kernels on real hardware, and the authors verified the
generated code and reviewed all text for accuracy.

\bibliographystyle{IEEEtranS}
\bibliography{reference}

\end{document}